\documentclass[pss,fleqn]{w-art}
\usepackage{times}
\usepackage{w-thm}
\usepackage{graphicx}
\begin{document}
\pagespan{1}{}



\title[Direct measurement of the electron-phonon relaxation rate in thin copper films]{Direct measurement of the electron-phonon relaxation rate in 
thin copper films}


\author[L.J. Taskinen]{L.J. Taskinen\footnote{Corresponding
     author: e-mail: {\sf lasse.taskinen@phys.jyu.fi}, Phone: +358 14 260 2359,
     Fax:+358 14 260 2351}\inst{1}} \address[\inst{1}]{NanoScience Center, Department of Physics, P.O. Box 35, 
FIN-40014 University of Jyv\"askyl\"a, Finland}

\author[J.M. Kivioja]{J.M. Kivioja\inst{2}} \address[\inst{2}]{Low Temperature Laboratory, Helsinki University of Technology, P.O. Box 2200, 02015 HUT 
Helsinki, Finland}
\author[J.T. Karvonen]{J.T. Karvonen\inst{1}}
\author[I.J. Maasilta]{I.J. Maasilta\inst{1}}
\maketitle

\begin{abstract}
We have used normal metal-insulator-superconductor (NIS) tunnel junction pairs, known as SINIS structures, for ultrasensitive thermometry at 
sub-Kelvin temperatures. With the help of these thermometers, we have developed an ac-technique to measure the electron-phonon (e-p) scattering rate 
directly, without any other material or geometry dependent parameters, based on overheating the electron gas. The technique is based on Joule heating 
the electrons in the frequency range DC-10 MHz, and measuring the electron temperature  in DC. Because of the nonlinearity of the electron-phonon 
coupling with respect to temperature, even the DC response will be affected, when the heating frequency reaches the natural cut-off determined by the 
e-p scattering rate. Results on thin Cu films show a $T^{4}$ behavior for the scattering rate,  in agreement with indirect measurement of similar 
samples and numerical modeling of the non-linear response.

\end{abstract}            
\renewcommand{\leftmark}
{L.J. Taskinen et al.: Direct measurement of the electron-phonon relaxation rate in thin copper films}

\section{Introduction}
\sloppy
Interaction between conduction electrons and thermal phonons is elementary for many processes and phenomena at low temperatures, and better 
measurements of the scattering rate need to be developed to characterise these processes accurately. Most earlier data was taken at DC, where sample 
parameters have to be taken into account before one can obtain the strength of the e-p interaction from a fitting parameter 
\cite{Roukes,well,kansk,Maasilta}. It would therefore be quite benefitial if one could measure the e-p scattering rate $1/\tau _{\rm e-p}$ directly 
without any fitting parameters and as a function of any external parameter. Some steps towards this direction have already been taken with a RF-based 
technique \cite{cle}.

Here we report a technique, which measures the e-p scattering rate directly without any fitting parameters and without the need of more complex RF 
circuitry. The rate is read directly from a cut-off frequency seen in the non-linear response of the DC electron temperature, measured with the help 
of NIS tunnel junctions at sub-Kelvin temperatures. 

In the disordered limit $ql<<1$, where $q$ is wavevector of the dominant thermal phonons and $l$ is the electron mean free path, theory predicts that 
the electron-phonon scattering rate from acoustic phonons is $1/\tau _{\rm e-p}=  \alpha T_e^4$ \cite{Schmid,Sergeev},
where $\alpha$ is a sample parameter and $T_e$ the electron temperature. In contrast, in the pure limit temperature dependence $1/\tau _{\rm e-p}=  
\alpha' T_e^3$ is expected \cite{Gantmakher}.   
\section{Experimental techniques}
\sloppy

\begin{figure}[htb]
\centering
\includegraphics[width=6.4cm, height=4.5cm]{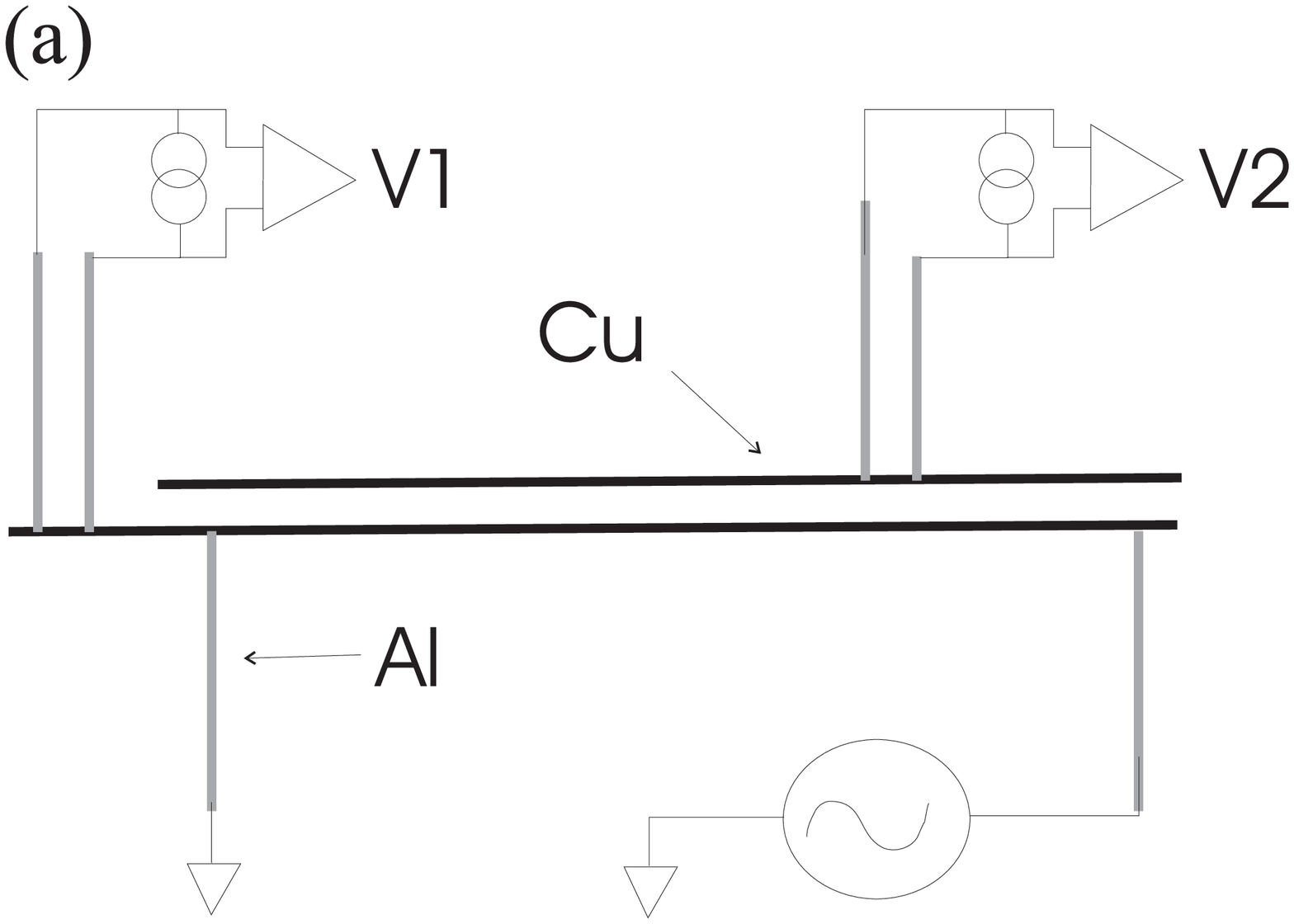} \includegraphics[width=6.4cm, height=4.5cm]{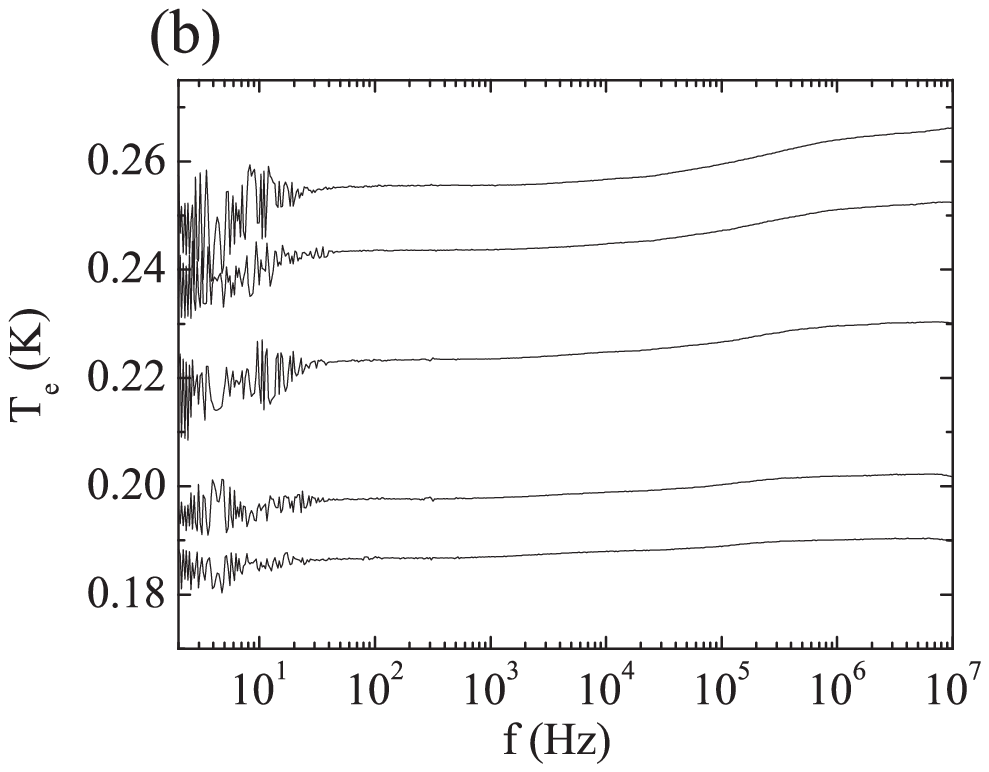}
\caption{(a) The measurement setup used. Black horizontal lines describe copper, and grey vertical lines aluminum. (b) The measured DC (average) 
electron temperature vs. heating frequency. Each line is an average of about 20  frequency sweeps.}
\label{fig:1}
\end{figure}

Our sample geometry is nearly identical to the one reported for DC measurements \cite{Maasilta,karv},  schematic of the sample and the measurement 
circuit is depicted in Fig. \ref{fig:1}(a). The sample used in this work has two Cu normal metal islands of length $\sim$ 500$\mu$m, width 300nm, and 
thickness 45nm, on oxidized Si substrate. The islands were separated by a distance of 2$\mu$m. The two aluminum wires going up in Fig. \ref{fig:1}(a) 
from each island were separated from Cu by a thin oxide layer, forming normal metal-insulator-superconductor (NIS) tunnel junctions in the overlap 
area. NIS junctions (SINIS structure) were used to measure the electron temperature of the Cu islands. Two aluminum wires going down from the lower Cu 
line were in metallic contact with the island, forming normal metal-superconductor (NS) junctions. These provided good electrical contact to the Cu, 
but the heat leak through them is negligible due to Andreev reflection, thus providing uniform heating power for the Cu wire. Heat leak through the 
tunnel junctions was also estimated to be at least an order of magnitude smaller than the cooling by electron-phonon coupling. 

SINIS thermometers were calibrated against a Ruthenium Oxide thermometer attached to the sample stage. The SINIS thermometers were current biased, and 
temperature of the cryostat (base temperature 60 mK) was changed slowly to keep the sample in thermal equilibrium with the sample stage, while the 
voltages of the SINIS thermometers were measured. There was a small discrepancy with the measured calibration curve and the corresponding BCS theory 
result at the lowest temperatures, but it can be explained by extra noise coming from the measurement circuitry to the sample \cite{Maasilta,karv}.

The experiment then proceeds by heating the electrons of the lower wire via the NS junctions with an ac-voltage. The frequency of the heating signal 
was changed step by step, and the voltages of both SINIS structures were measured simultaneously. The upper Cu island in Fig. \ref{fig:1}(a) could be 
used to measure the phonon temperature \cite{Maasilta,karv}, whereas the lower SINIS measured the heated electron temperature. We took many frequency 
sweeps with the same ac-voltage amplitude, and averaged the data to reduce the noise. The amplitude of the ac-voltage was then changed to obtain a 
different average electron temperature. A typical sweep consisted of 300-500 frequency points between 0.1Hz-5MHz. The five representative sweeps shown 
in Fig. \ref{fig:1}(b) are an average of approximately 20 frequency sweeps each. Overheating is clearly seen in the high frequency range between 
10kHz-1MHz for all the sweeps shown, which can be understood by the numerical modeling discussed below.

\section{Numerical modeling}

To fully understand the effect of ac heating at a signal frequency $\omega$ on the electron temperature $T_e$, we need to solve the full differential 
equation governing the heat flow:

\begin{equation}
T_e\frac{dT_e}{dt}= -A(T_e^n-T_p^n)+P'[1-\cos(2\omega t)],
\label{diffeq}
\end{equation}

where we have defined $A=\Sigma'/\gamma$ and $P'=P_{ac}/(2\gamma \Omega)$, where $\Sigma'$ is the sample constant in the electron-phonon interaction 
power $P=\Sigma'\Omega(T_e^n-T_p^n)$, $\Omega$ is the sample volume, $\gamma$ is the Sommerfeld constant and $P_{ac}$ is the amplitude of the ac 
power. Since $\Sigma'$ also depends linearly on $\gamma$, we point out that the temperature relaxation rate does not depend on the heat capacity at 
all, and is only dependent on the electron-phonon scattering rate. 

\begin{figure}[htb]
\centering
\includegraphics[width=16cm, height=6.5cm]{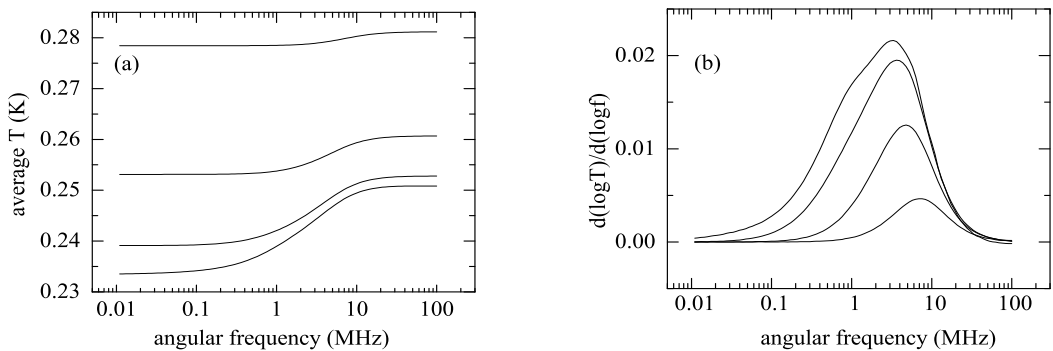}
\caption{(a) Simulated average electron temperature. (b) Logarithmic derivative of the simulation data.}
\label{fig:num1}
\end{figure}

We have integrated Eq. (\ref{diffeq}) numerically for exponents $n=5,6$, corresponding to $1/\tau _{\rm e-p}= \alpha T_e^m$, $m=3,4$. From the 
$T_e(t)$ curve, the steady state average electron temperature has been determined as a function of $\omega$, with varying ratio of ac heating $P_{ac}$ 
to DC heating power $\Sigma'\Omega T_p^n$, and with realistic sample parameters. Fig. \ref{fig:num1}(a) shows the obtained results for four different 
power ratios in the case $n=6$ (the impure limit). As can be seen, the average $T_e$ develops a clear step up at some cut-off frequency, and this 
frequency moves up as a function of $T_e$. To be able to define this cut-off unequivocally, we analyse the data further by taking the logarithmic 
derivative with respect to $\omega$, shown in Fig. \ref{fig:num1}(b). The log-derivative develops a clear peak, whose position we will use as the 
definition of the cut-off frequency $\omega_c$. In addition, the log-derivative shows quite clearly the effect of the strength of the non-linearity: 
the larger the relative ac power (top curve largest), the more non-symmetric the peak is. 

\begin{figure}[htb]
\centering
\includegraphics[width=7cm, height=5cm]{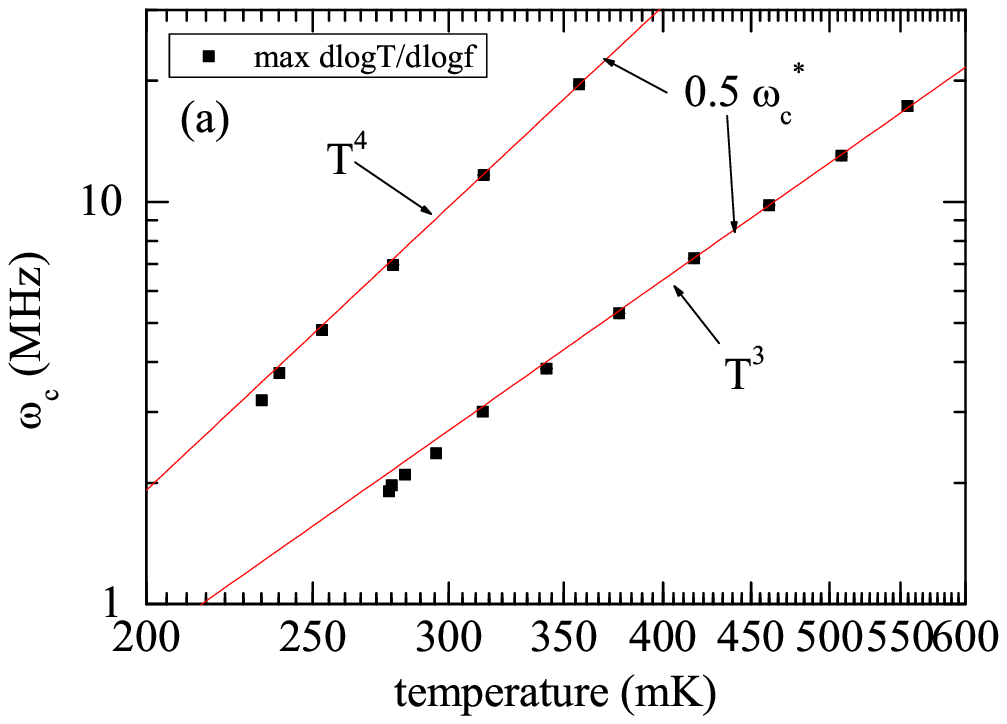}\includegraphics[width=6.5cm, height=4.5cm]{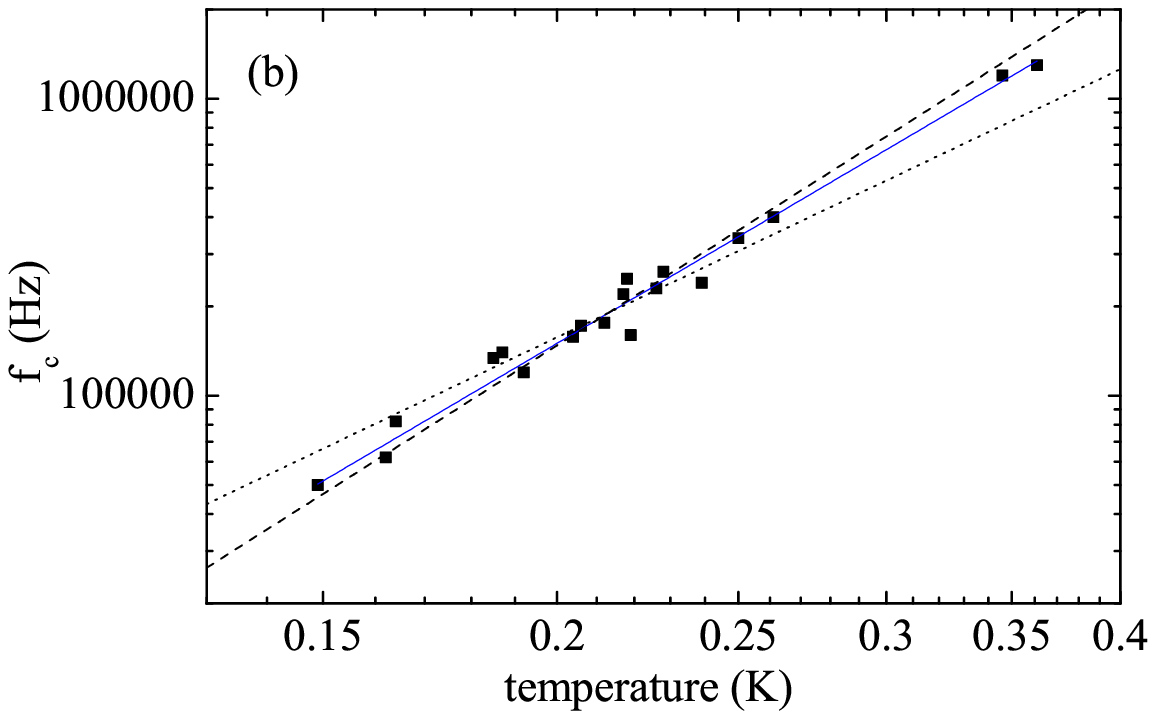}
\caption{(a) Maximum of the logarithmic derivative of the simulated data (squares) vs. $T_e$. $\omega_{c}^{*}=nAT_{e}^{n-2}$. (b) Maximum of the 
logarithmic derivative obtained from the measured data (squares). Dotted line correspond to $ f_c \propto T^3$ and dashed to $ f_c \propto T^4$. Solid 
line is a best fit $f_c=aT^n$, with $n=3.7$. }
\label{fig:num2}
\end{figure}

If one plots $\omega _c$ vs. the low-frequency limit of $T_e$, one obtains the plot shown in Fig. \ref{fig:num2}(a), where both cases $n=5$ and $n=6$ 
are shown. It is quite clear that $\omega _c$ follows the right temperature dependence expected for a cut-off determined by the e-p scattering rate 
only, as long as the nonlinearity is not too severe (the lowest temperatures deviate a bit in this case). 
Therefore, we can state with confidence that the analysis scheme outlined above will be a direct measurement of the e-p scattering rate without any 
fitting parameters.

\section{Experimental results}

Figure \ref{fig:num2}(b) shows the experimental data analyzed as outlined above, with the maxima of the log-derivatives of the measured data vs. the 
low-frequency limit of $T_e$.  The best two parameter fit $f_c=aT^n$ gives $n=3.7$. However, the scatter in the data is still a bit high, and to make 
a more qualitative analysis, we also compare the integer powers $n=3$ and $n=4$. As can be seen,  the dashed line corresponding to power $T^4$ fits to 
data much better than the one corresponding to $T^3$. This temperature dependence is in agreement with the DC measurements in the disordered regime 
\cite{Maasilta, karv}. From the $T^4$ fit we get $\Sigma'=1.0\times 10^{10}$W/K$^6$m$^3$, using a literature value $\gamma=0.69\frac{mJ}{mol*K^{2}}$ 
for copper, again in agreement with our DC data. One should keep in mind, however, that the numerical value of $\Sigma'$ is strongly dependent on the 
sample disorder. In fact, we have seen in other, very similar samples that the value of $\Sigma'$ can be almost an order of magnitude smaller. 
\section{Conclusions}
\sloppy
We have presented a method to measure the electron-phonon scattering rate directly without the need to know any material or geometry dependent sample 
parameters. The measurements can be performed at DC, greatly simplifying the electronics, due to the non-linearity of the e-p interaction. Our initial 
experimental results agree with the DC data for Cu thin films in the disordered limit.

\begin{acknowledgement}
 We thank D.-V. Anghel, J. P. Pekola and A. Sergeev for discussions.  This work was supported by the Academy of Finland under the Finnish Center of 
Excellence Program 2000-2005 (Project No. 44875), and by the Academy of Finland project No. 105258. L.J.T acknowledges partial financial support by 
Vilho, Yrj\"o and Kalle V\"ais\"al\"a foundation.
\end{acknowledgement}


\end{document}